\documentclass[%
 aip,
 amsmath,amssymb,
 reprint,%
]{revtex4-1}

\usepackage{graphicx}
\usepackage{dcolumn}
\usepackage{bm}

\usepackage{float}
\usepackage{algorithm}
\usepackage{algpseudocode}
\usepackage{caption}
\usepackage{tikz}          
\usepackage{subcaption}    
\usetikzlibrary{positioning} 

\usepackage[utf8]{inputenc}
\usepackage[T1]{fontenc}
\usepackage{mathptmx}
\usepackage{etoolbox}
\usepackage{graphicx}
\usepackage{booktabs}
\usepackage{mathrsfs}

\makeatletter
\def\@email#1#2{%
 \endgroup
 \patchcmd{\titleblock@produce}
  {\frontmatter@RRAPformat}
  {\frontmatter@RRAPformat{\produce@RRAP{*#1\href{mailto:#2}{#2}}}\frontmatter@RRAPformat}
  {}{}
}%
\makeatother
\begin{document}


\preprint{AIP/123-QED}

\title{Turbulence enhancement of a fan array wind generator using geometric texturing and optimization-based control}

\author{Gengshou Cao}
  \affiliation{School of Robotics and  Advanced Manufacturing, Harbin Institute of Technology, Shenzhen 518055, People's Republic of China}

\author{Tamir Shaqarin}
\affiliation{Department of Mechanical Engineering, Tafila Technical University, Tafila, 66110, Jordan}

\author{Zhutao Jiang}
  \affiliation{School of Robotics and  Advanced Manufacturing, Harbin Institute of Technology, Shenzhen 518055, People's Republic of China}

\author{Yutong Liu}
  \affiliation{School of Intelligence Science and Engineering, Harbin Institute of Technology, Shenzhen 518055, People's Republic of China}

\author{Yiqing Li }
  \affiliation{Department of Mechanical Engineering, University College London, London WC1E 7JE, UK}

\author{Nan Gao}
\affiliation{Hanghua Technologies, Inc., Dalian 116023, People's Republic of China}

\author{Xiaozhou He }%
 \email{Contact author: hexiaozhou@hit.edu.cn}
  \affiliation{School of Robotics and  Advanced Manufacturing, Harbin Institute of Technology, Shenzhen 518055, Peoples' Republic of China}
  \affiliation{Low-Altitude Economy Science and Technology Research Institute, Harbin Institute of Technology Shenzhen, Shenzhen 518055, People's Republic of China}

\author{Bernd R.\ Noack}%
 \email{Contact author: bernd.noack@szu.edu.cn}
  \affiliation{College of Mechatronics and Control Engineering, Shenzhen University, Canghai campus, Shenzhen 518060, People's Republic of China}

\date{\today}

\begin{abstract}
Fan array wind generators (FAWG) are designed to generate a rich set of turbulent flows reminiscent of those found in natural environments. In this study, we experimentally investigate a square FAWG consisting of $10\times10$ individually controllable fans with 4 cm width and a maximum velocity of 17 m/s. The goal is to maximize the turbulence intensity in the test region. Two approaches for fan operation are investigated: first, geometric texturing of the duty cycle distribution, and second, maximization of the turbulence intensity at selected hot-wire sensors with particle-swarm optimization. We find that geometric texturing (specifically a checkerboard pattern) yields a robust, uniform turbulence field (Tu $\approx$ 0.14) driven by jet interactions. Conversely, particle swarm optimization achieves higher local turbulence (Tu $\approx$ 0.28) but significantly sacrifices spatial uniformity. This study underscores the trade-off between local maximization and global uniformity in active turbulence generation.
\end{abstract}

\maketitle

\section{Introduction}

The rapid development of Unmanned Aerial Vehicles (UAVs) and electric Vertical Take-Off and Landing (eVTOL) aircraft is transforming military, civilian, and urban transportation sectors~\cite{hassija2021fast, li2020autonomous}. As these vehicles are increasingly deployed in complex urban environments, their stability and safety are critically challenged by atmospheric disturbances, particularly high-intensity wind turbulence~\cite{zhao2023fluid, alsamhi2019survey}. Velocity fluctuations induced by such turbulence can lead to severe position deviations and compromised flight path reliability~\cite{khan2022emerging}. To certify the airworthiness of these next-generation aircraft, it is essential to replicate these hazardous conditions in a controlled laboratory setting.

Traditional wind tunnels typically rely on passive techniques to generate turbulence, such as grids, roughness elements, or spires~\cite{hurst2007scalings}. While effective for creating standard boundary layers, these passive methods suffer from bulky designs and a lack of flexibility; they cannot rapidly switch between flow regimes or generate the extreme, non-stationary turbulence required for rigorous UAV testing~\cite{coppola2009experimental, greenblatt2016unsteady}.

To address these limitations, Fan Array Wind Generators (FAWG) have emerged as a versatile alternative. By utilizing a matrix of individually controllable fans, FAWGs enable the precise modulation of velocity profiles and the generation of active turbulence. The concept was pioneered by Nishi et al.~\cite{nishi1997turbulence}, who demonstrated that independent fan speed control and oscillating blades could actively regulate turbulence intensity and Reynolds stresses. Building on this foundation, Cao and Cao~\cite{cao2017toward} designed a 120-fan array capable of reproducing longitudinal and vertical fluctuations. More recently, Kikuchi et al.~\cite{kikuchi2019low} introduced cost-effective architectures using PC cooling fans, democratizing the technology for smaller research facilities.

Recent research has deepened our understanding of FAWG aerodynamics. Li et al.~\cite{li2024aerodynamic} characterized the flow physics of small-scale arrays, revealing that the near-field is dominated by swirling fan-blade interactions that evolve into jet-like structures. Ozono et al.~\cite{ozono2006turbulence} proposed an ``active grid mode,'' activating fans in alternating grid patterns to generate isotropic turbulence with favorable spectral properties. Most recently, Liu et al.~\cite{liu2025aerodynamic} extended these findings to large-scale facilities ($40\times40$ fans), confirming that the fundamental flow laws—such as the potential core contraction—scale predictably from small to large arrays.

Despite these advances, the majority of FAWG literature has focused on reproducing mean wind profiles or standard atmospheric boundary layer (ABL) statistics. Few studies have systematically explored control strategies specifically designed to \textit{maximize} turbulence intensity for extreme limit testing.

In this study, we address this gap by investigating two distinct strategies to maximize turbulence intensity in a $10\times10$ FAWG: (1) Geometric Texturing Patterns (GTP), which utilize discrete checkerboard actuation to induce massive shear layer mixing, and (2) Machine Learning (ML) Optimization, specifically Particle Swarm Optimization (PSO)~\cite{shaqarin2023fast}, driven by real-time hot-wire feedback. We characterize the resulting flow fields using hot-wire anemometry and Particle Image Velocimetry (PIV). Unlike prior work that aims for smooth ABL replication, this research provides a framework for generating high-intensity, challenging turbulent environments tailored for UAV stability testing.

The remainder of this manuscript is organised as follows. Section II details the experimental facility, control methods, and measurement techniques used in this study. Section III presents a comparative study of the flow fields resulting from geometric patterns and machine-learning optimized states, and elucidates the governing physical mechanisms. Finally, Section IV summarises the key findings and provides guidelines for future FAWG turbulence design.

\section{METHODOLOGY}
\label{sec:2}

This section describes the experimental setup and turbulence optimization. 
This test facility is used to study the turbulence level generated from fan arrays 
with symmetric on-off patterns of fans
and or machine learning optimized fan control. 
In Sec.\ \ref{sec:2.1},  we present a detailed description of the fan array wind generator. 
In Sec. \ref{sec:2.2}, we  introduce the turbulence optimization methods.
Sec. \ref{sec:2.3} details the 
flow measurement employed in this investigation. 
All physical quantities are listed in Table~\ref{tab:qua}.

\begin{table}
\caption{\label{tab:qua} 
Table of physical quantities.}
\begin{ruledtabular}
\begin{tabular}{lcr}
Physical quantities & Symbols & Units \\ 
\hline
Turbulence intensity & $Tu$ & \text{---} \\
The downstream distance of the wind  & $x$ & mm \\
Measurement spatial resolution & $\Delta y, \Delta z $ & mm \\
Time & $t$ & s \\
Time-averaged streamwise velocity component & $U$ & m/s \\
Streamwise velocity component & $u$ & m/s \\
Streamwise fluctuation velocity component & $u'$ & m/s \\
Width of 10 $\times$ 10 FAWG & $W$ & 40 cm \\ 
Width of a single fan element & $w$ & 4 cm \\ 
Environmental temperature & $T$ & ℃ \\ 
\end{tabular}
\end{ruledtabular}
\end{table}

\subsection{Fan-Array Wind Generator}
\label{sec:2.1}
\begin{figure}[hbt!]
    \centering
    \includegraphics[width=0.4\textwidth]{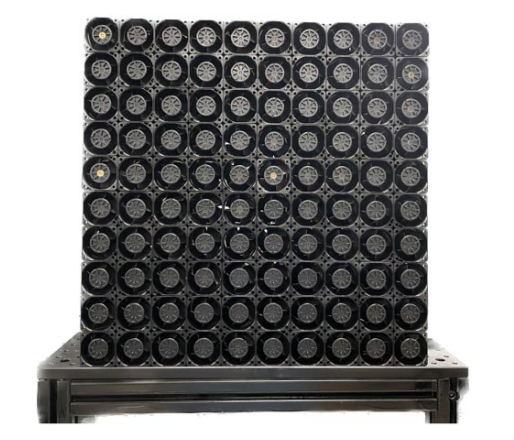}
    \caption{ Fan array wind
generator}
    \label{fig:FAWG}
\end{figure}

As illustrated in Fig~\ref{fig:FAWG}, the fan array wind generator system comprises 10$\times$10 matrix of axial-flow fan units. Each unit employs a two-stage counter-rotating fan (model GFB0412EHS, Delta Electronics) with a cubic configuration measuring 40$\times$40$\times$56 mm. The front-stage blades rotate counterclockwise, while the rear-stage blades operate in the clockwise direction.

The mechanical assembly integrates 100 fan units arranged in a 10 $\times$ 10 array. 
The fans are mounted in frames and multi-hole washers.  
The frame is mounted on a 1.5-meter-tall aluminum extrusion stand.
The system operates with a 12V DC power supply. Speed regulation of individual fan units is achieved via a multi-channel control architecture utilizing seven PCA9685 pulse width modulation (PWM) driver modules. 
Each module generates 16 independent PWM signals. The duty cycle of all 100 PWM signals are commanded via the Arduino Mega microcontroller to control the rotational speed of the fans. 
While the system inherently supports spatio-temporally variable airflow generation,
this study only investigates steady duty cycles corresponding to steady fan wind velocities.

\subsection{Actuation of the fan array}
\label{sec:2.2}

\begin{figure}[hbt!]
    \centering
    \includegraphics[width = 0.35\textwidth]{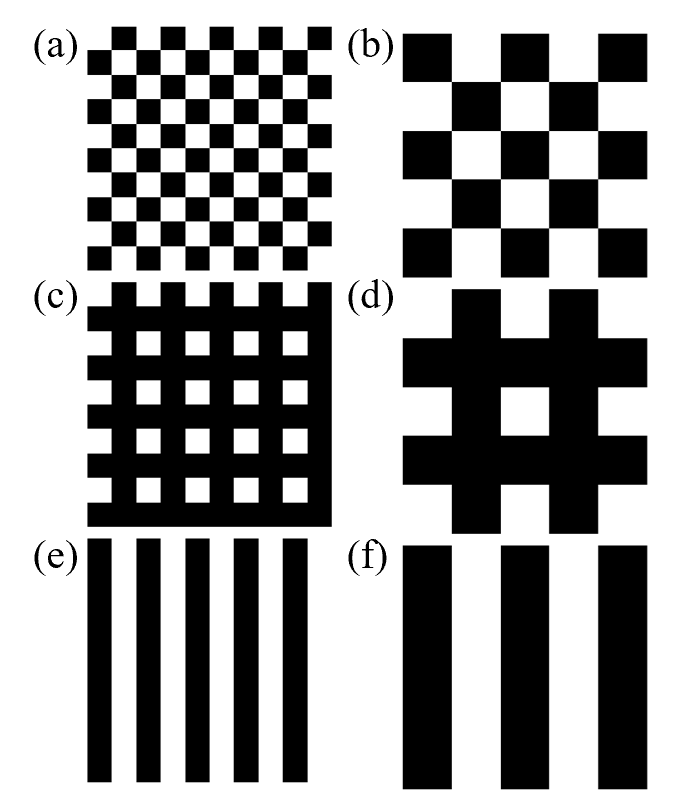}
    \caption{Geometric texturing pattern:Black: active; White: inactive.}
    \label{fig:physics-inspired pattern}
\end{figure}

The turbulence intensity
is first investigated with symmetric patterns of on-off states. 
 Fig~\ref{fig:physics-inspired pattern} illustrates regions with activated fans (operating at 100\% power, represented in black) and deactivated fans (off state, represented in white). 
 Chess board, 
 vertical rows
 and grid-like actuation patterns are explored for different sizes.
 These discrete aerodynamic surface patterns
 shall be referred to geometric texturing pattern (GTP).
 The boundaries between active and non-active areas generate a shear layer leading to fluctuations evolving in streamwise direction. 

A larger actuation space is explored 
with machine learning methods.
Now, 
all fans can assume arbitrary velocities between zero and the maximum value.
An up-down right-left symmetric actuation pattern is assumed limiting the number of indepedend control commands from 100 to 25.
A fast‑converging particle swarm optimization through targeted,
position‑mutated, elitism (PSO‑TPME) is used
after good experience with analytical test functions\cite{shaqarin2023fast}
and high-dimensional turbulence control experiments\cite{Shaqarin2024sr}.
The duty cycle of the PWM is taken as actuation command. This duty cycle is nearly linearly related to the fan array velocity.
The PSO-TPME algorithm is detailed in Appendix A. 

Fig~\ref{fig:flow chart} illustrates the whole control system for optimizing the parameters of the fan array. 
Here, $\bm{b}$ represents the 25 control parameters, 
and $J$ is the cost function
to be minimized. 
For each $\bm{b}$ generated by the optimizer, the corresponding control parameters are transmitted to the controller. 
The hot wires measure the flow velocity at the target points and determine the cost function which is fed back to the optimizer to generate the next set of test control commands $\bm{b}$.

Thus, the problem of maximizing turbulence intensity is reformulated as an optimization problem, with the goal of finding the optimal parameters $\mathbf{b} = [ d_{ij} ]$ 
to minimize the cost function $J$.
Here, $d_{ij}$ denotes the duty cycle of the PWM control signal for the fan located in the $i$-th row and $j$-th column of the $10 \times 10$ fan array, 
where $d_{ij} \in [0, 100]$.
Only the first $5 \times 5$ duty cycles
need to be optimized as the pattern is symmetrically continued:
$$ d_{ij} =  d_{11-i,j} =  d_{i,11-j} 
= 
 d_{11-i,11-j} \forall i, j \in \{1,\ldots 5\} $$

 The goal is to maximize the averaged turbulence intensity
\begin{equation}
  {\rm Tu} = \frac{1}{4} \sum_{i=1}^4  \frac{u_{\rm rms,i }}{U_i}
  \end{equation}
 where $U_i$ denotes the mean velocity measured with  the $i$th hot-write probe and
  $u_{{\rm rms},i}$ the corresponding root-mean-square fluctuation level.
 In addition, the measured mean flow
 shall be close to the target velocity 
 $U_{\rm target}$
 and fluctuations shall be as uniform as possible. 

This optimization goal is served by minimizing the reciprocal of the turbulence intensity 
and penalizing the deviation from the mean target velocity as well as difference between the turbulence intensities.
The chosen corresponding cost function reads
\begin{equation}
J = J_a + J_b + J_c
\label{eq:J_total}
\end{equation},
where 
$J_a = 1/{\rm Tu}$ quantifies
the overall fluctuation level,
$J_b = \frac{\lambda_1}{4} \sum_{i = 1}^{4}\left|\frac{U_{i}}{U_{\text{target}}}-1\right|$
the deviation between the local mean flows
from the target velocity
and $J_c = \frac{\lambda_2}{4} \sum_{i = 1}^{4}\left|\frac{{\rm Tu}_{i}}{{\rm Tu}}-1\right|$
the difference between the turbulence itensities ${\rm Tu}_i$ measured by the $i$th hot-wire probe.
The penalization factors $\lambda_1$ and $\lambda_2$ are choosen so that one percent increase in turbulence intensity is approximately one percent deviation from 
the target velocity and fluctuation uniformity.
These values are obtained converging experience in numerous experiments.
The resulting optimized parameters ideally find the global minimum,
\begin{equation}
\label{eq:J}
\bm{b^{*}} = \arg \min_{\bm{b}\in \mathscr{R}^{25}} J(\bm{b})
\end{equation}

\begin{figure}[hbt!]
    \centering
    \includegraphics[width=0.5\textwidth]{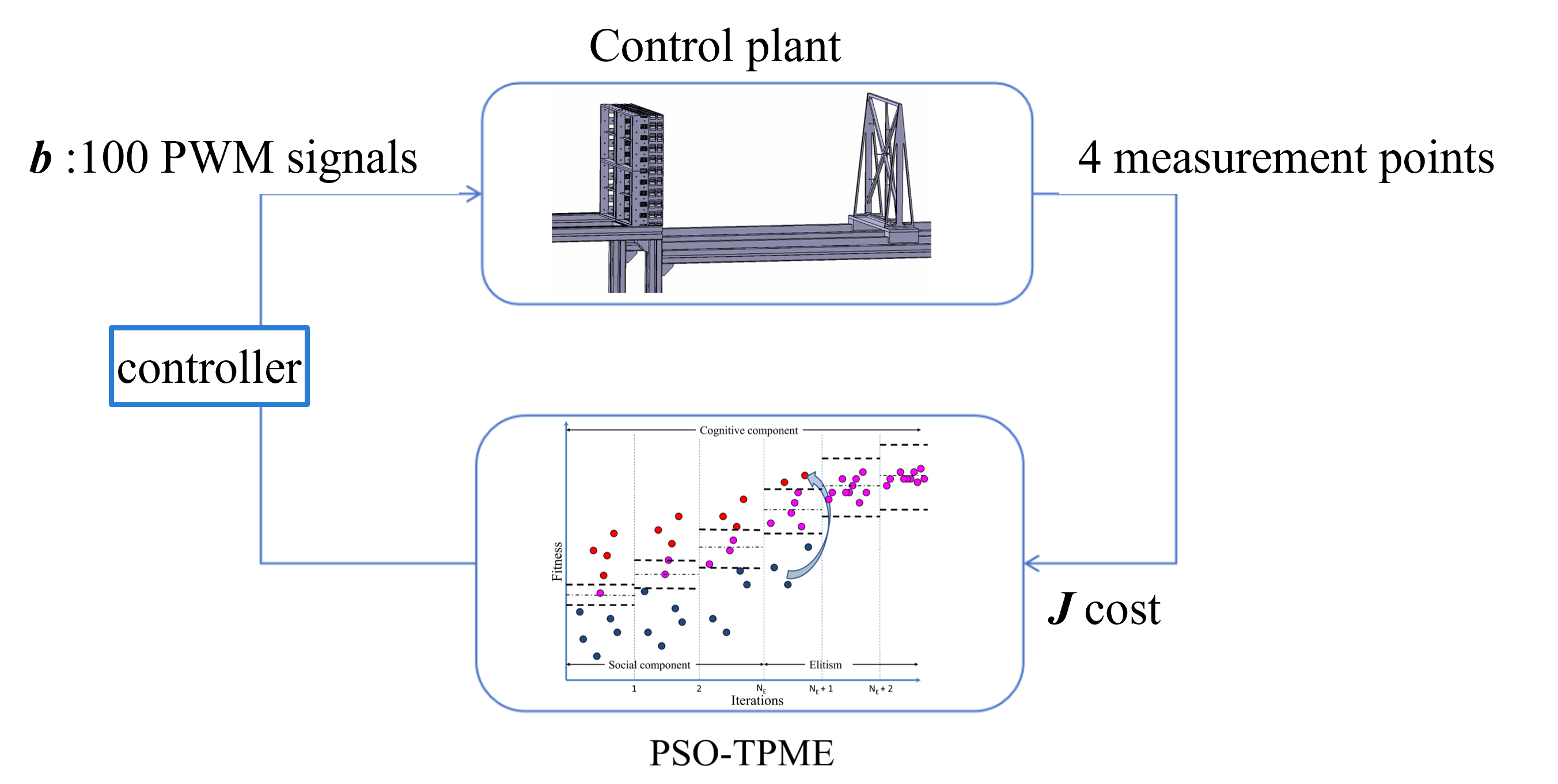}
    \caption{Principal sketch of the optimization-based control.}
    \label{fig:flow chart}
\end{figure}

\subsection{Hot-Wire Anemometry and Particle Image Velocimetry}
\label{sec:2.3}

\begin{figure}[hbt!]
    \centering
    \includegraphics[width=0.5\textwidth]{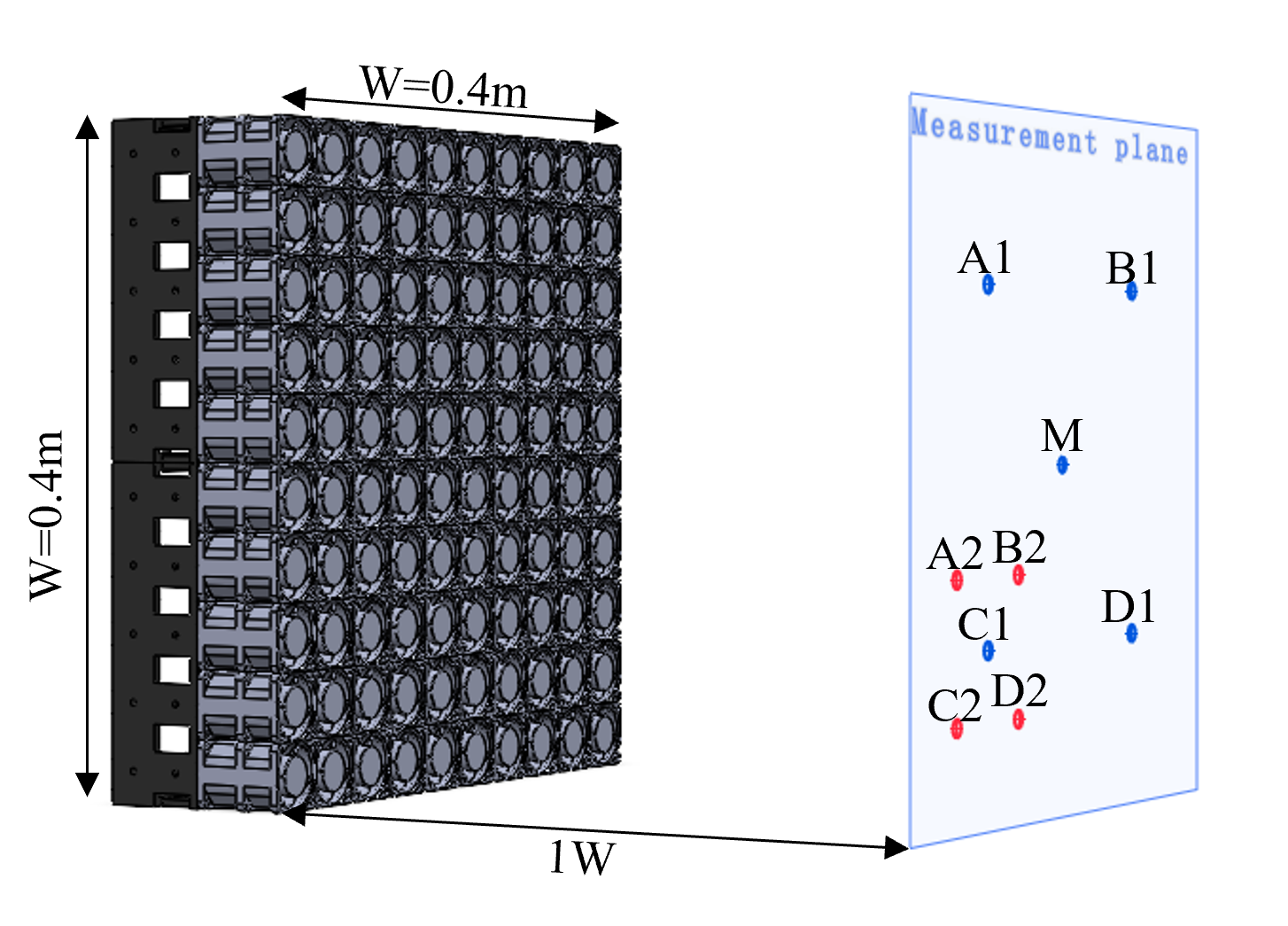}
    \caption{Visualization of hot-wire anemometer measurement points. The width of the 10$\times$10 fan-array is denoted by W}
    \label{fig:target-points-position}
\end{figure}
In this study, 
the flow is analyzed with hot-wire anemometry (HWA) and particle image velocimetry (PIV) is employed. 
The flow is described in 
a Cartesian coordinate system $(x,y,z)$ with its origin at the center of the fan-array exit plane 
displayed in Fig~\ref{fig:target-points-position}. 
Here, 
$x$ denotes the streamwise direction, while $y$ and $z$ represent the horizontal and vertical directions, respectively.
The characteristic dimensions 
are the height $h$ of the individual fan elements and $H$ for the entire array.

\begin{table*}
\caption{\label{tab:setup}Flow field measurement setup}
\begin{ruledtabular}
\begin{tabular}{l l l}
Experiment & Measurement planes/points & Data sampling \\ 
\hline
HWA &
M, A1, A2, B1, B2, C1, C2, D1, and D2\\
& at $x/w = 1,\ldots,20$ &
$F_s = 10$~kHz, $t = 30$~s \\ 
2D3C PIV & $x/w = 1$ & 2000 snapshots \\
\end{tabular}
\end{ruledtabular}
\end{table*}

Table~\ref{tab:setup} summarizes the experimental campaigns conducted in this study.
The hot-wire anemometer (Hanghua CTA04) is used to measure the streamwise velocity component at $x/w = 1,...,20$.
The selected cross-plane measurement points are shown in Fig.~\ref{fig:target-points-position}, where points A1, B1, C1, and D1 are positioned at the midpoints of the four quadrants defined by the y- and z-axes, while points A2, B2, C2, and D2 are clustered in the lower-left quadrant.
Probes were rigidly mounted on a modular traversing system ensuring positional accuracy $\pm 0.5$ mm.
The HWA calibration employed a jet calibrator (0--30 m/s) referenced to a Pitot tube.
The fluctuating voltage output was then digitized using a National Instruments USB-6009 14-bit data acquisition device. 
A fourth-order polynomial is fitted to map the relationship between the voltage outputs and the corresponding velocities.
In addition, this constant-temperature anemometer is  compensated to achieve optimal dynamic performance. 
Active temperature control maintained ambient conditions at $25 \pm 2^{\circ}\text{C}$ despite heat loads from full-power FAWG operation (100\% duty cycle). The accuracy of hot-wire measurements and the repeatability of experiments are discussed in Appendix B.

Particle image velocimetry (PIV) is employed to characterize the velocity field on the cross-stream ($y$-$z$) plane at $x/H=1$ in the fan array far field, with the measurement plane visualized in Fig~\ref{fig:target-points-position}.  
In this study, we employ a LaVison PIV system equipped with a double-pulsed laser and two 2752 $\times$ 2200 megapixel charge-coupled device cameras. 
For far-field measurements, a stereoscopic setup is used to calculate the three-dimensional velocity vectors. 
In this case, two cameras with Scheimpflug adapters are mounted on opposite sides of the wind generator. 
The calibration of the camera system is achieved by detecting and fitting the target points on a three-dimensional calibration plate via a pinhole model.  
A dual-cavity Litron Nano L 200-15 Nd:YAG laser with a 532 nm wavelength illuminates seeding particles from an Antari fog machine that produces particles about 0.2 $\mu$m in size. 
PIV image pairs are acquired at a frequency of 12 Hz under three representative modes: geometric texturing pattern, machine learning derived pattern, and uniform pattern. 
In each mode, 2000 snapshots are recorded for stereoscopic PIV measurements on each far-field plane.

\section{RESULTS AND DISCUSSION}
\label{sec:3}

This section compares the flow characteristics produced by the geometric texturing pattern and the PSO-TPME-driven pattern. To assess the proposed approaches' capabilities to maximize turbulence intensity, we use FAWG with a $10 \times 10$ fan grid and a $2 \times 2$ hot-wire grid, as discussed in Sec. ~\ref{sec:2.3}.

\subsection{Geometric texturing pattern}
\label{sec:subsection1}

\begin{figure*}[ht!]
    \centering
    \includegraphics[width=0.9\textwidth]{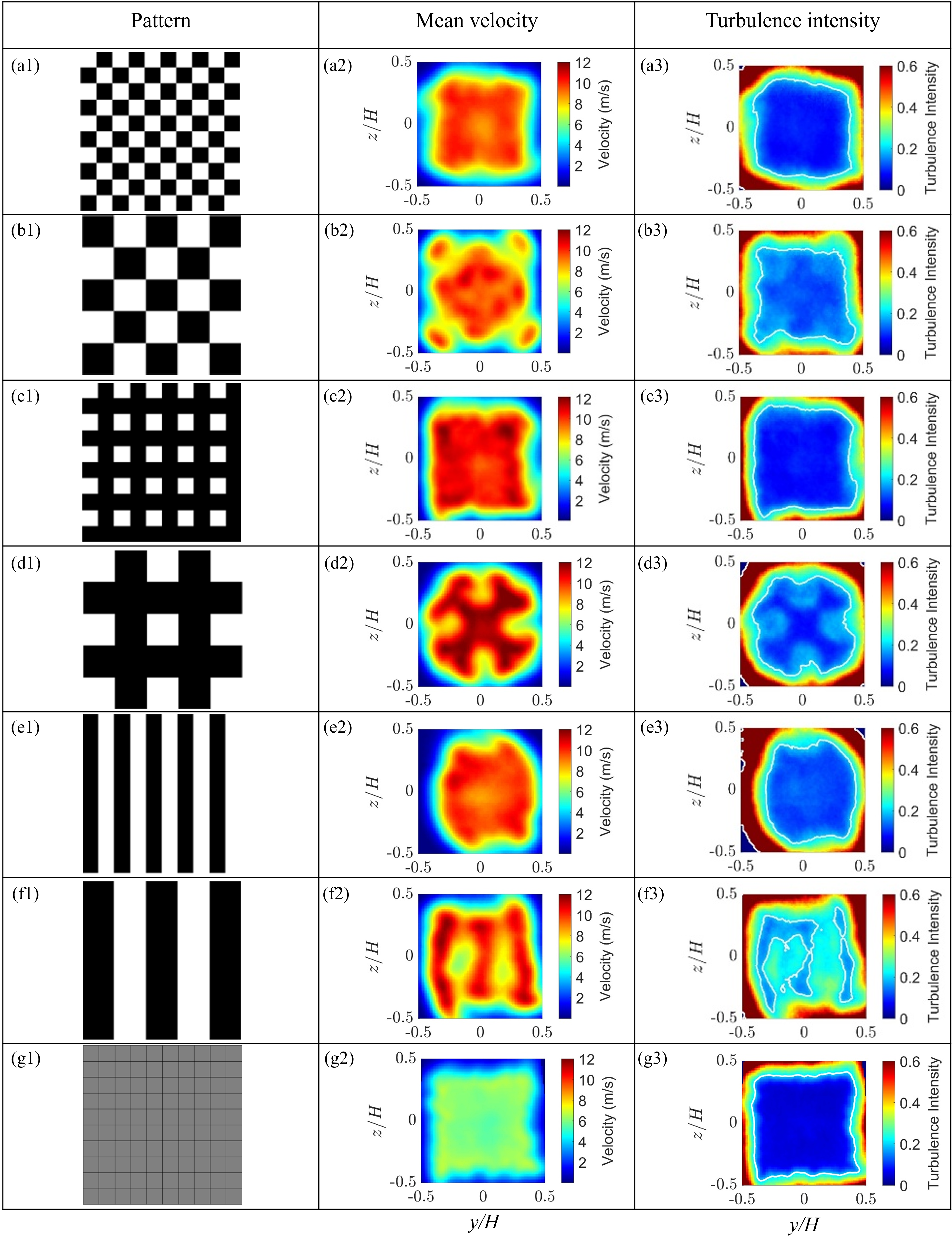}
    \caption{PIV results of the geometric texturing patterns. The first column illustrates the fan activation configurations: black denotes fully activated fans, white represents inactive (powered-off) fans, and gray indicates fans operating at 50\% power. The second column shows the mean velocity fields measured in the PIV plane, while the third column presents the corresponding turbulence intensity distribution. In all color maps, blue indicates lower values and red indicates higher values.}
    \label{Fig:table-model-PIV}
\end{figure*}

Fig~\ref{Fig:table-model-PIV} presents the measured streamwise velocity and turbulence intensity distributions downstream of six different geometric texturing patterns. Significant variations in both the magnitude and spatial uniformity of the flow properties can be observed as the fan activation pattern changes.

For the small checkerboard pattern (Pattern 1), the mean velocity field exhibits a highly uniform distribution, forming a smooth circular, relatively high-speed core with only a narrow low-speed region near the boundaries. The turbulence intensity remains low to moderate, and its spatial distribution is remarkably uniform, as indicated by the smooth, concentric low turbulence intensity contour. Similarly, a comparable level of uniformity appears in the large checkerboard pattern (Pattern 2), although mild petal-shaped modulations in the velocity field introduce slightly enhanced spatial variability while preserving overall symmetry. The velocity field irregularities are reflected in the turbulence intensity field, which demonstrates slight distortions compared to Pattern 1.

The small cross-stripe pattern (Pattern 3) yields the highest local mean velocity magnitude, reaching up to 12 m/s, while preserving high uniformity in both velocity and turbulence intensity. In contrast, the large cross-stripe pattern (Pattern 4) produces pronounced non-uniformities, with ring-like features in the turbulence intensity and a four-lobed pattern in the velocity distribution. Next, the narrow vertical stripe pattern (Pattern 5) generates a uniform field comparable to Pattern 1, with a slight vertical elongation consistent with the stripe orientation. The turbulence intensity remains low in the center but shows a similar vertically elongated distribution. The wide vertical stripe pattern (Pattern 6) leads to the strongest spatial anisotropy, producing alternating high- and low-speed bands and correspondingly uneven turbulence intensity.

Pattern 1 and Pattern 2 stand out for their exceptional yet complementary characteristics: Pattern 1 yields a highly uniform turbulence intensity field, while Pattern 2 produces the highest turbulence levels. Given these distinctive advantages—and their clear pattern correlation—both modes are selected for further analysis.

\begin{figure*}[ht!]
    \centering
    \includegraphics[width=1\textwidth]{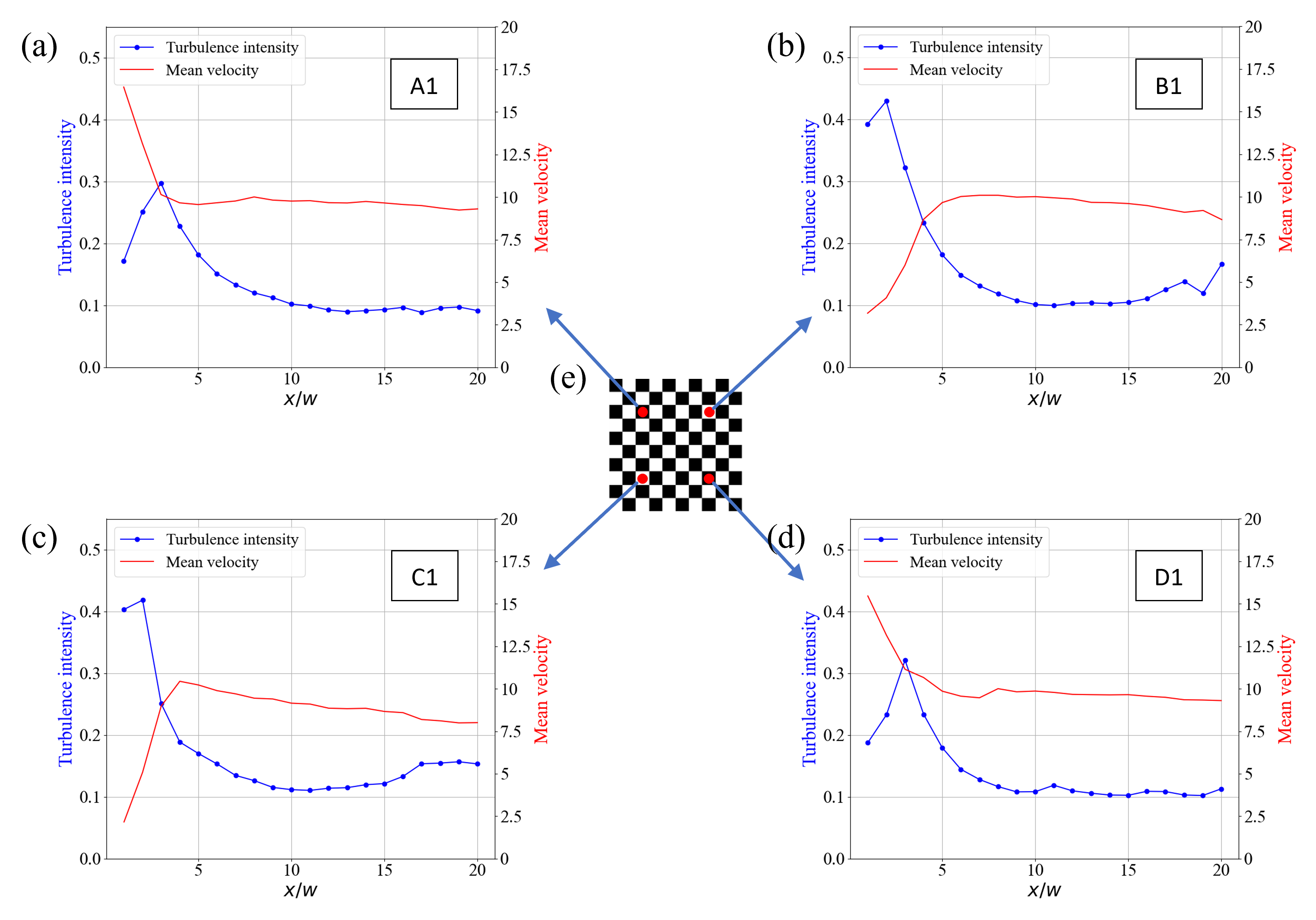}
    \caption{Streamwise evolution of the flow generated by Pattern 1. The time-averaged streamwise velocity (red lines) and turbulence intensity (blue lines) of points $A1$, $B1$, $C1$, and $D1$ are depicted in sub-figures (a), (b), (c), and (d), respectively, with their exact locations shown in (e). The HWA is obtained across twenty cross-stream planes ($x/w= 1...,20$).}
    \label{Fig:small-big-points}
\end{figure*}

\begin{figure*}[ht!]
    \centering
    \includegraphics[width=1\textwidth]{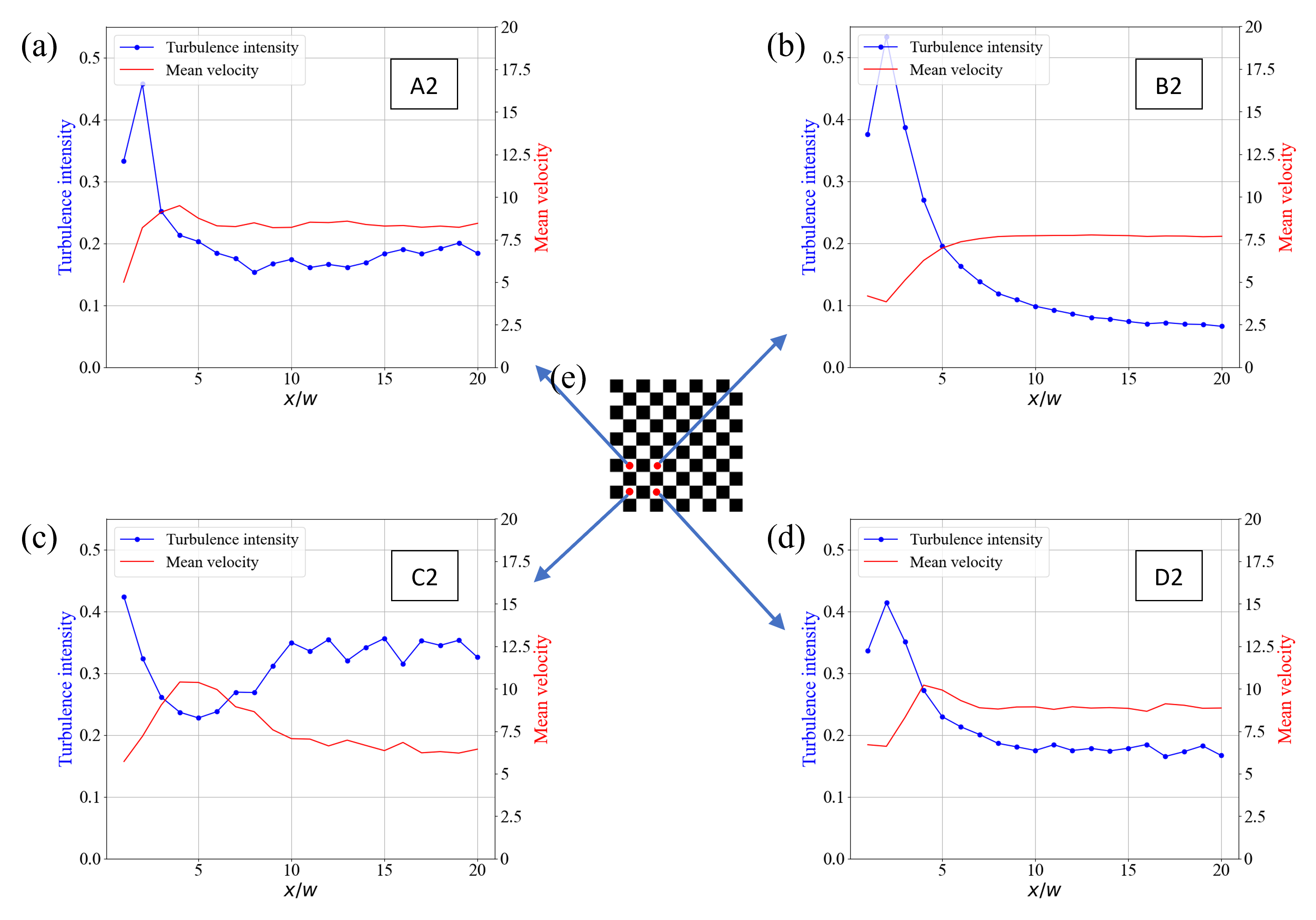}
    \caption{Streamwise evolution of the flow generated by Pattern 1. The time-averaged streamwise velocity (red lines) and turbulence intensity (blue lines) of points $A2$, $B2$, $C2$, and $D2$ are depicted in sub-figures (a), (b), (c), and (d), respectively, with their exact locations shown in (e). The HWA is obtained across twenty cross-stream planes ($x/w= 1...,20$).}
    \label{Fig:small-small-points}
\end{figure*}

As depicted in Fig~\ref{Fig:small-big-points}, the flow development downstream of four symmetrical points in the small checkerboard pattern shows that the individual fully-activated fan (black square region) exhibits classical jet behavior. 
Around $2.5$ small-fan widths or checkerboard cell sizes downstream ($x/w = 2.5$), the flow field exhibits peak turbulence intensity ${\rm Tu} = 0.355 \pm 0.065$. 
Farther downstream, at approximately $x/w = 10$, the mean velocity and turbulence intensity of the flow field both gradually become more uniform  as the jets interact and merge. Fig~\ref{Fig:small-small-points} also shows the same downstream evolution. Ultimately, the flow becomes fully developed, with a mean velocity of $8.65 \pm 0.65$ m/s and a turbulence intensity around ${Tu} = 0.125 \pm 0.035$.

As shown in Fig~\ref{Fig:small-small-points}, we selected four measurement points within one quadrant of the small checkerboard pattern  to enhance the spatial resolution of the flow field characterization. 
With the origin as the reference center, sub-figures (a) and (d) display similar velocity evolution at the same radial distances. This strong similarity verifies that the airflow produced by the cooperative operation of multiple fans exhibits the behavior of interacting jets.
Notably, the measurement point associated with sub-figure (c) is located  closest to the outer boundary of the measurement domain, positioning it near the jet shear layer. Consequently, unlike the monotonic decay  observed at other locations, the turbulence intensity at this location remains elevated at approximately 0.35. 
This behavior shows high consistency with the findings on turbulence characteristics in jet boundary layers reported ~\citet{li2024aerodynamic} using Particle Image Velocimetry (PIV), further reinforces the reliability of the present experimental results.

With all fans operating at maximum rotational speed, the average streamwise velocity reaches approximately 12 m/s at $x/w = 10$.
 In contrast, Fig.~\ref{Fig:small-big-points} and~\ref{Fig:small-small-points} reveal that activating only half of FAWG at 100\% PWM produces a mean velocity reaches 8.5 m/s—only 70.8\% of the fully activated case, and significantly lower than the proportional theoretical scaling.
However, the maximum wind speed generated by a single fan can reach 17.5 m/s, this speed is not sustained in dense array configuration. 
This discrepancy may be attributed to the axial-flow fan used in the present study, which is more susceptible to variations in system resistance compared to centrifugal fans.
When fans are arranged into a closely packed array, it becomes difficult to simultaneously satisfy the optimal inlet and outlet conditions for all individual units. The dense fan array interference and flow interactions reduce the flow rate at the array level.

\begin{figure*}[ht!]
    \centering
    \includegraphics[width=1\textwidth]{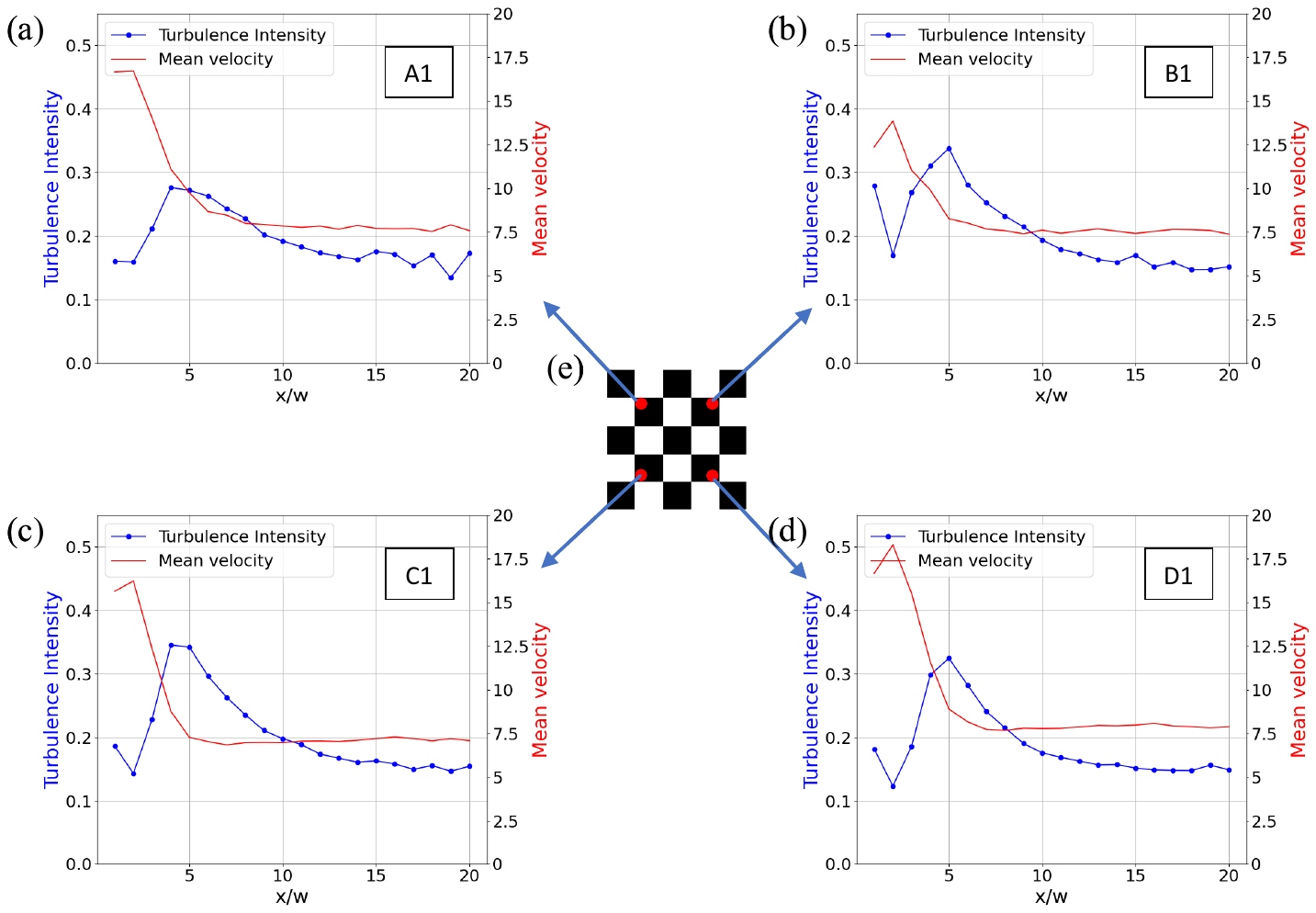}
    \caption{Streamwise evolution of the flow generated by Pattern 2. The time-averaged streamwise velocity (red lines) and turbulence intensity (blue lines) of points $A2$, $B2$, $C2$, and $D2$ are depicted in sub-figures (a), (b), (c), and (d), respectively, with their exact locations shown in (e). The HWA is obtained across twenty cross-stream planes ($x/w= 1...,20$).}
    \label{Fig:big-big-points}
\end{figure*}

\begin{figure*}[htb!]
    \centering
    \includegraphics[width=1\textwidth]{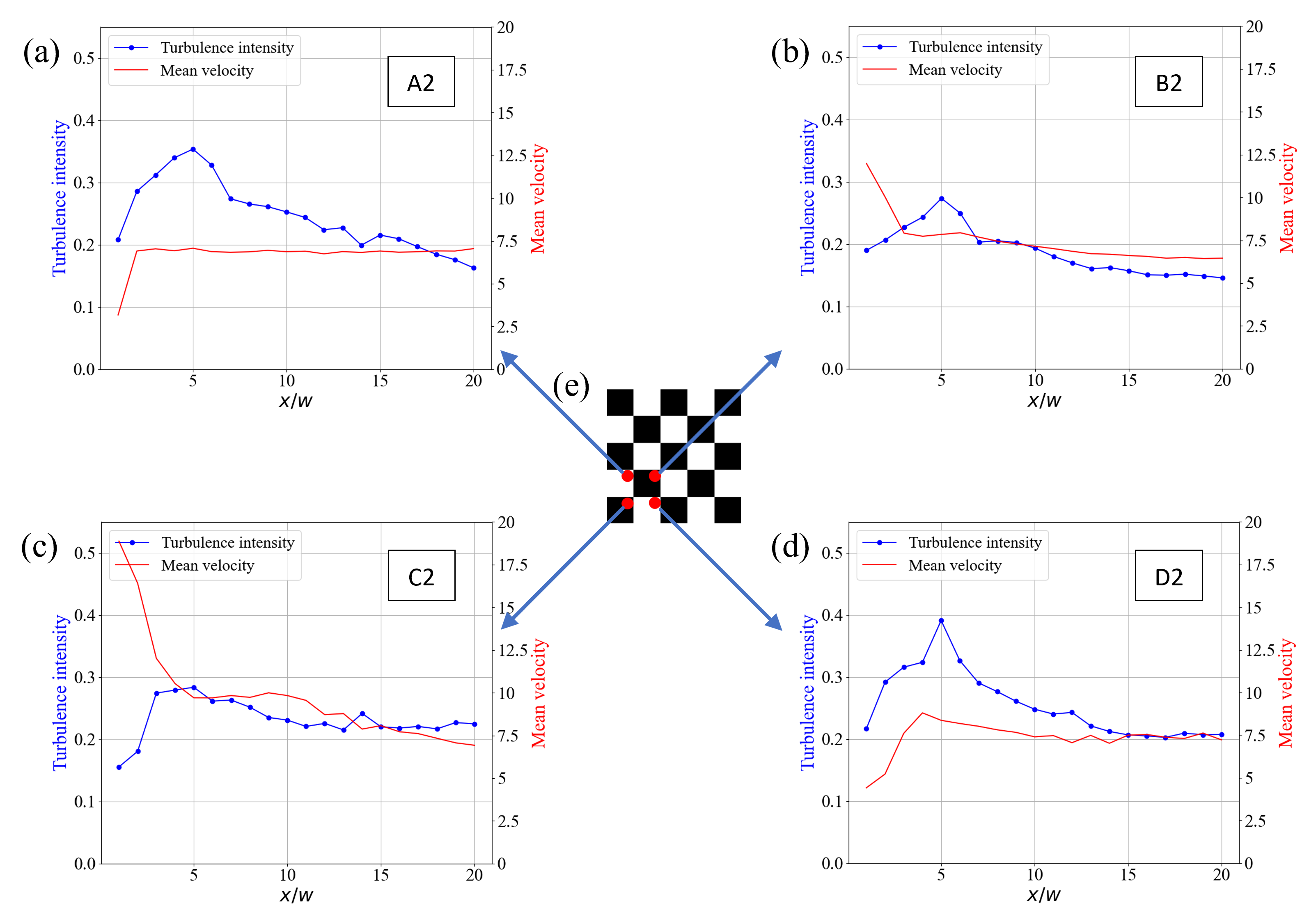}
    \caption{Streamwise evolution of the flow generated by Pattern 2. The time-averaged streamwise velocity (red lines) and turbulence intensity (blue lines) of points $A2$, $B2$, $C2$, and $D2$ are depicted in sub-figures (a), (b), (c), and (d), respectively, with their exact locations shown in (e). The HWA is obtained across twenty cross-stream planes ($x/w= 1...,20$).}
    \label{Fig:big-small-points}
\end{figure*}

Fig~\ref{Fig:big-big-points} shows that for the large checkerboard pattern, the flow enters the mixing-dominated region at approximately $x/w = 5$, corresponding to a distance of 2.5 checkerboard cell size downstream. 
This transition is characterized by a peak in turbulence intensity of approximately 30\%.
The same behavior is observed in Fig.~\ref{Fig:big-small-points}. The higher maximum turbulence intensity at other points in Fig~\ref{Fig:big-small-points} may be due to the proximity of the measurement points to the boundary layer.
Finally, the fully-developed flow reaches a mean velocity of $7.7 \pm 0.2$ m/s with turbulence intensity ${Tu} = 0.15 \pm 0.02$.
Spatial evolution analysis further reveals that velocity in activated (black) regions decay downstream, whereas velocities in the deactivated (white) regions increase due to pressure-driven entrainment.
The mixing zone length scale $L_m$ is found to scale linearly with the characteristic checkerboard cell dimension $h$: $L_m / h = 2.5$, validating the geometric similitude principle for jet-array interactions.

Additionally, we observe in Fig~\ref{Fig:big-big-points} that the turbulence intensity decreases immediately after $x/w=1$ before rising again, while the mean velocity continues to increases. However, this trend is not fully evident in Fig~\ref{Fig:big-big-points}, suggesting possible spatial variability in the early development region. We hypothesize that in this location (half of the checkerboard cell size downstream), the airflow is still in the initial mixing stage, where the shear layer is not fully developed. This initial mixing stage is completed within one characteristic length scale (i.e., the checkerboard cell size).

As illustrated in Fig~\ref{Fig:big-small-points}, we compared and analyzed the flow field characteristics of two checkerboard control patterns using PIV measurements (Fig~\ref{Fig:table-model-PIV}(a) and Fig~\ref{Fig:table-model-PIV}(b)).
The large checkerboard pattern (employing a 2$\times$2 fan unit as the control unit) generates a distinctive "one dominant, four secondary" velocity distribution structure, accompanied by a significant increase in turbulence intensity within the central region.
Within the core span of the flow field, defined as 
$\Omega_c = \bigl\{ (x,z) \mid x, z \in [-120, 120]\,\mathrm{mm} \bigr\}$, 
the average turbulence intensity reached $0.14$, indicating that enlarging the control unit size effectively enhances flow mixing.
In contrast, the small checkerboard pattern (using a single fan as the control unit) produces a more spatially uniform velocity distribution. However, the average turbulence intensity in its core region reached only 0.09. This clear discrepancy demonstrates a significant difference in flow control mechanisms between the two patterns.
Overall, these findings provide crucial experimental evidence for optimizing turbulence control strategies in multi-fan systems.

Comparative experiments reveal that the geometric texturing pattern significantly impacts wind-field characteristics.
Control experiments (Fig~\ref{Fig:table-model-PIV}a and~\ref{Fig:table-model-PIV}g), conducted with all 10$\times$10 fans operating uniformly at 50\% power, reveal two key findings:
(1) Residual flow distortion persists even under uniform operation mode. This observation confirms that flow asymmetry primarily stems from the inherent characteristics of fans rather than from the control strategy.
(2) The small checkerboard mode demonstrates superior turbulence enhancement, exhibiting significantly higher turbulence intensity compared to the uniform mode.\\
Specifically, the average turbulence intensity within the core region is 80\% higher than the corresponding value of 0.05 measured in the uniform mode.
This finding underscores the advantages of discrete spatial actuation for turbulence control and suggests novel approaches for active flow control in wind tunnels.

\subsection{PSO-TPME Optimization-Based Control Pattern}
\label{sec:subsection2}


The loss function of the optimization algorithm comprises three aspects: turbulence intensity at the target points, velocity uniformity at the target points, and uniformity of turbulence intensity at the target points. In this work, the weighting coefficients of the loss function $\lambda_1$ and $\lambda_2$ were both set to 5.5.

\begin{figure}[h]
    \centering
    \includegraphics[width=0.5\textwidth]{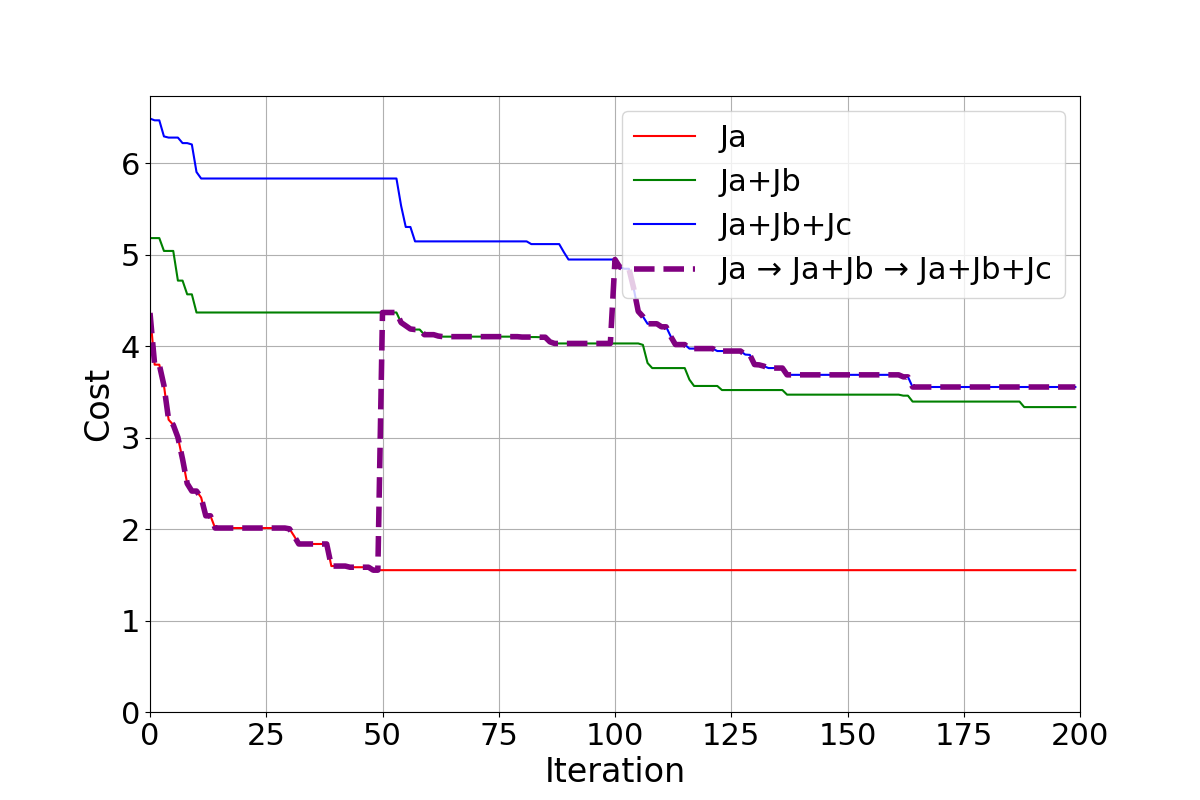}
    \caption{Learning curve of the FAWG maximum turbulence intensity}
    \label{Fig:learning_curve}
\end{figure}

The optimization learning curve is shown in Fig~\ref{Fig:learning_curve}.
 The purple dashed line indicates that the optimization objective is divided into three stages. Initially, the optimizer targets only the turbulence intensity at four prescribed locations, allowing the optimizer to quickly identify patterns with high turbulence intensity, but without yet focusing on the mean flow velocity. Subsequently, both the turbulence intensity and the uniformity of velocity are taken into account that provides more control authority on the flow velocity and turbulence intensity. Finally, the objective function is expanded to include the spatial deviation of the turbulence intensity along with the previous factors, enabling convergence toward a fully integrated objective function for FAWG. The red curve represents the evolution of the reciprocal of the turbulence intensity as iterations progress. The green curve illustrates the progressive improvement in velocity uniformity across iterations, while the blue curve tracks the optimization of the turbulence uniformity metric. Collectively, these curves illustrate the effectiveness of multi-stage optimization that steers the optimization process from merely generating turbulence to a more controlled and spatially uniform turbulence generation problem.

\begin{figure}[hbt!]
    \centering
    \includegraphics[width=0.5\textwidth]{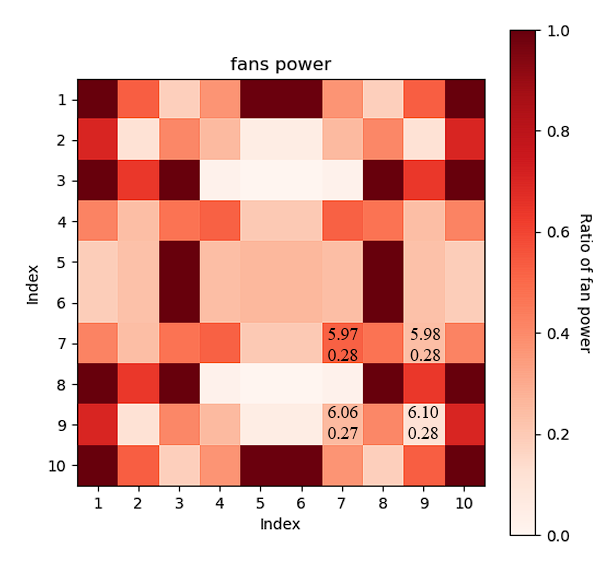}
    \caption{Optimized fan-array actuation pattern obtained using online PSO-TPME. }
    \label{Fig:FANSPOWER}
\end{figure}

Fig.~\ref{Fig:FANSPOWER} shows the optimized PWM input distribution of the $10\times10$ fan. The measurement locations are shown in the lower right quadrant, where the upper value is the measured speed, and the lower value is the turbulence intensity for each sensor location. As shown in the figure, the optimizer successfully tracks the reference velocity (6$m/s$) with less than $\pm2\%$ error, and the optimizer also maximizes the turbulence intensity to 28\% with spatial uniformity error within $\pm4\%$. However, the optimizer in this setup has no access to velocity and turbulence from other locations, as it only relies on the sensors from the lower right quadrant. Therefore, there is no authority from this control strategy on the flow characteristics in other quadrants.
For instance, the figure shows that the fans located in the central two rows and two columns were assigned relatively lower PWM input values. The resulting effects on the velocity field and turbulence intensity map can not be detected using the current measurement system.
Because of the limited number of hot-wire sensors used in this work, the measurement system cannot capture the characteristics of the entire flow field. This limitation motivates the use of PIV measurements to obtain the entire flow field information. 


\begin{figure}[hbt!]
    \centering
    \includegraphics[width=0.3\textwidth]{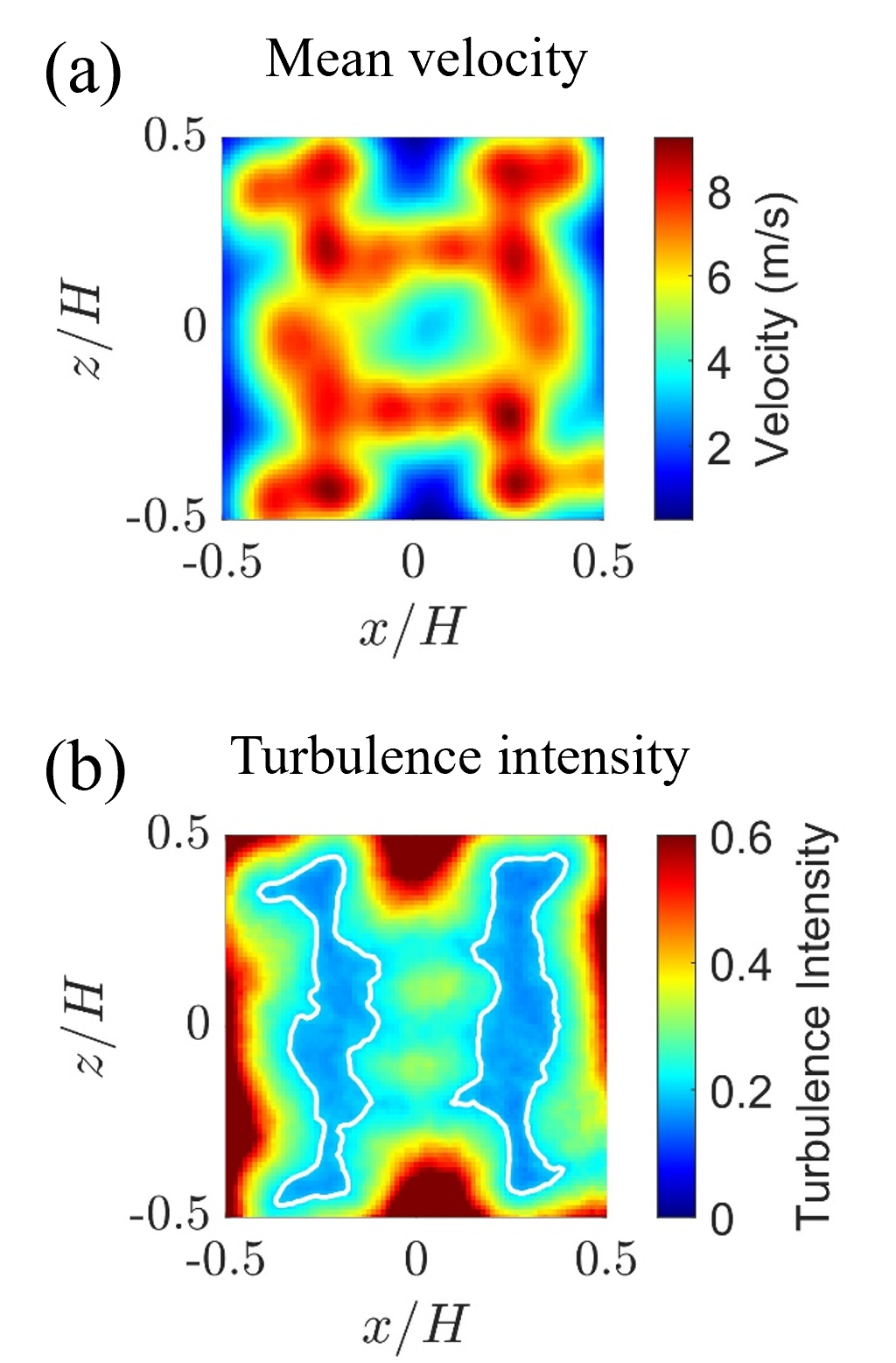}
    \caption{Cross-plane PIV measurement results of the optimal result from the optimization: (a) time-averaged velocity field distribution, (b) turbulence intensity distribution.}
    \label{Fig: PIV-best}
\end{figure}

The optimized velocity and turbulence intensity distributions from the PIV measurements are shown in Fig~\ref{Fig: PIV-best}. The white contour lines in Fig~\ref{Fig: PIV-best}(b) indicate the 20\% turbulence intensity boundary. Although the velocity at the target point accurately reached 6 $m/s$ (verified by hot-wire measurements), the full-field PIV results reveal that the overall velocity distribution lacks uniformity, with significant velocity differences between measurement points. For example, the figure shows that the velocity and turbulence intensity in the central part of the entire array are both relatively low. In other words, the optimization algorithm prioritizes the target points at the expense of the other points. This indicates that the four hot-wire sensors do not effectively represent the turbulence intensity of a larger area. The parameter optimization method using mirror replication performs particularly poorly at the edges of the array.
Nevertheless, based on the turbulence intensity map, the proposed method performed well, with the average turbulence intensity in the core region of the flow field reaching 0.23, which is much higher than that of the checkerboard mode. This highlights the capability of optimization-based control to enhance turbulence generation. Future research can further optimize the uniformity of the flow field by increasing the number and spatial distribution of the target measurement points.

\section{Conclusions}
\label{sec:4}

This study experimentally investigates a 10$\times$10 fan-array wind generator (FAWG) composed of individually controllable fans.
The objective is maximizing turbulence intensity in the whole test region
while keeping a nearly uniform mean velocity.
Focus is placed on a typical drone testing location 
one width downstream.
The velocity of each actuator is treated 
as free steady control variable.
The optimization goal is achieved 
with particle swarm optimization (PSO)
and with an exploration of physics inspired on-off patterns (textures) of the fan array velocities
monitored by hot-wire sensors.
All results can be interpreted in terms of the unsteadiness
of  shear-layers emanating from the fan-array wind generators.
The proposed results guide future efforts
towards simultaneous design of wind profiles 
and turbulence intensities.

The current study extends uniform blowing results
by the authors for this small fan-array wind generator  \cite{li2024aerodynamic} 
and the world's largest FAWG
in terms of number of fans
build with 40 $\times$ 40 fans with 8 cm width  \cite{LiuYutong2025pf,LiuYutong2025guinness}.
For both FAWG, the near-field flow can be characterized as a jet with nearly uniform `potential core' surrounded
by a shear with large fluctuation level.
The transversal extent of the uniform core decreases nearly linearly within 5 FAWG widths to zero.

Following specific conclusions can be drawn for the interrogation plane one FAWG width downstream.
First, for uniform blowing, 
the core features a turbulence intensity of 0.05. 
These fluctuations can be traced back from the highly unsteady and nonuniform exit flows from the fans.

Second, the fluctuation level can be increased 
with symmetric geometric patterns of on-off fan activity.
Differently sized textures with active fans are explored:
a checker-board pattern, vertical stripes and a grid-like pattern.
The largest turbulence level of 0.14 is obtained
with a coarse checker-board activation pattern.  
The size of the active squares is one-fifth of the downstream distance.
In other words, these large fluctuations arise at the end of the `potential core' of the small actuation jets.
Here, the largest fluctuations of individual jets are typically observed.
The turbulence intensity slowly decrease in streamwise direction
and significantly abover 0.1 in the first two width.
Consistent with this interpretation, 
smaller fluctuation levels are observed 
for fine-grained actuation patterns.
These results indicate that the best size of active areas
for the maximum fluctuation level is one-fifth of the streamwise distance.

Third, fluctuation levels can be further increased using machine learning optimizers and allowing for arbitrary fan velocities between 0 and the maximum values.  A right-left and top-down symmetry is enforced 
in the actuation pattern to monitor the flow at 4 $\times$ 4
interrogation points covering nearly uniformly 
the `potential core region' with only 4 hot-wire sensors.
PSO-TPME \cite{shaqarin2023fast} yields
a uniform mean flow with a turbulence level of 0.28,
i.e.\ twice as large as the texture-based actuation.
These large fluctuation levels are achieved 
by large inactive fan array regions leading 
to much lower velocities and 
much lower fluctuation levels between the sensors.
In other words, PSO-TPME over-optimizes 
the turbulence level at the sensor location 
and a much finer sensor resolution is required
for a more uniform turbulence generation.

In summary, the maximum achievable turbulence level
with fan array wind generators 
using an optimized steady, nonuniform  fan blowing 
seems to be the maximum value 
for shear layers between vanishing and uniform flow.
Machine learning optimizers require a dense array of hot-wires
to prevent unrepresentative local values.
Investigated on-off fan array textures have, as expected, 
lower turbulence intensities 
which peak for active areas with a one-fifth of the downstream width.
Future efforts of the authors are directed towards
commanding both mean-flow profiles and turbulence intensity.
We anticipate that mean flow can be effectively controlled with kernel-based distributed input distributed output control \cite{Shaqarin2025pf}
while the fluctuation level is controlled with textures of on-off patterns within realizable limits.

\appendix
\section{PSO-TPME}
\label{acl}

\begin{figure*}[hbt!]
    \centering
    \includegraphics[width=1\textwidth]{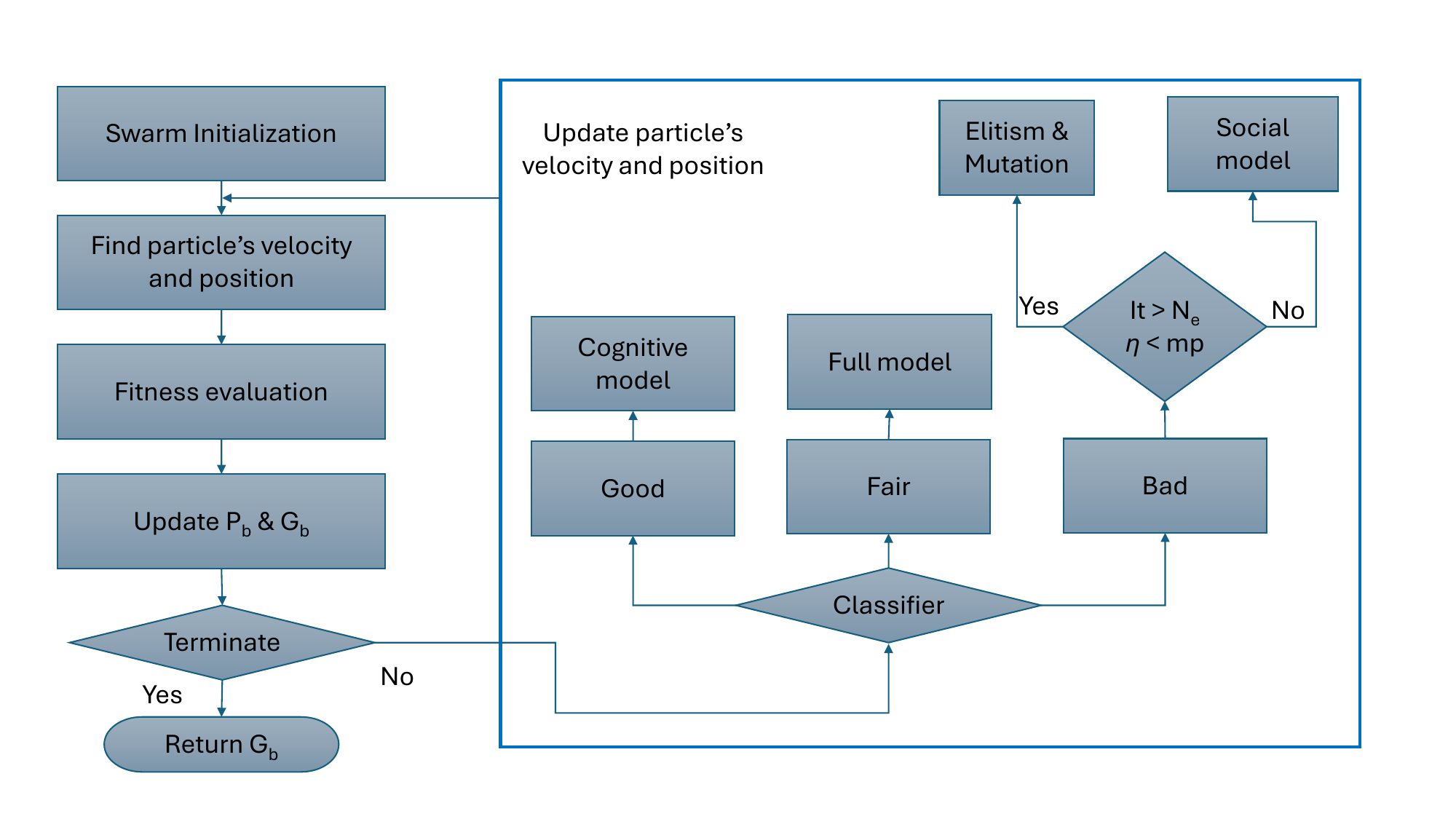}
    \caption{PSO-TPME algorithm flowchart (\citet{shaqarin2023fast})}
    \label{fig:PSO}
\end{figure*}

Recently, \citet{shaqarin2023fast} proposed particle swarm optimization with targeted position mutated elitism (PSO-TPME), an enhanced particle swarm optimization (PSO) variant with improved capabilities regarding both convergence speed and global exploration. PSO-TPME has demonstrated fast, real-time performance for optimization-based control \cite{shaqarin2023particle, Shaqarin2024sr}. PSO-TPME (Algorithm~\ref{alg:cap}) integrates three cooperative mechanisms: adaptive classification, elitism, and targeted position-mutation (see Fig~\ref{fig:PSO}), which are outlined below.
\subsection{Adaptive Fitness-Based Auto-terminated Classifier}
The swarm's mean fitness $\bar{F}$ is calculated every iteration. A classification range is then defined using a prescribed classification percentage $\rho$.  The classifier then categorizes the particles based on their individual fitness $F_i$ into three groups:

\paragraph{Good particles:}
\begin{equation}
F_i < (1-\rho)\,\bar{F}
\end{equation}
Their velocities are reduced to strengthen exploitation by relying on the cognitive component.

\paragraph{Fair particles:}
\begin{equation}
(1-\rho)\,\bar{F} \le F_i \le (1+\rho)\,\bar{F}
\end{equation}
They are updated using both cognitive and social components to maintain a balance between exploration and exploitation.

\paragraph{bad particles:}
\begin{equation}
F_i > (1+\rho)\,\bar{F}
\end{equation}
Their velocities are updated primarily through the social component to enhance exploration.

This adaptive classification regularly reforms the swarm, accelerating convergence while maintaining diversity in the whole swarm.
\subsection{Elitism and Targeted Position-Mutation }

Particles that stay in the bad class for more than $N_e$ consecutive iterations are labeled \emph{hopeless particles}, implying they fail to progress to a higher class despite full exploration freedom.

PSO-TPME replaces these hopeless particles by directly relocating them toward the available global-best position. This elitism step rapidly improves the swarm's exploitation capability and redirects computational effort toward better regions of the search space. However, elitism alone risks early stagnation at a local minimum due to the loss of diversity. Consequently, PSO-TPME imposes a targeted position-mutation in the vicinity of the global-best position, which will diversify the swarm and improve the exploration capability of the optimizer.
With mutation probability $m_p$, each hopeless particle receives a mutated version of the global-best solution:

\begin{equation}
\mathbf{x}_i \leftarrow 
\mathbf{g} \left(2\alpha \eta + (1-\alpha)\right),
\end{equation}
where $\eta \in [0,1]$ is a uniform random number and $\alpha$ is the mutation range. Overall, the classification, elitism, and mutation features yield a swarm that vigorously converges while maintaining high exploration capabilities.

\begin{algorithm}[H]
\caption{PSO-TPME}
\label{alg:cap}
\begin{algorithmic}[1]

\State Initialize swarm positions $\mathbf{x}_i$ and velocities $\mathbf{v}_i$
\State Initialize personal bests $\mathbf{p}_i$ and global best $\mathbf{g}$

\For{$k = 1$ to $K_{\max}$}
  \State Compute swarm mean fitness $\bar{F}$

  \For{each particle $i$}
    \If{$F_i < (1-\rho)\bar{F}$}
      \State $\mathbf{v}_i \gets \omega \mathbf{v}_i
      + c_1 r_1 (\mathbf{p}_i - \mathbf{x}_i)$

    \ElsIf{$(1-\rho)\bar{F} \le F_i \le (1+\rho)\bar{F}$}
      \State $\mathbf{v}_i \gets \omega \mathbf{v}_i
      + c_1 r_1 (\mathbf{p}_i - \mathbf{x}_i)
      + c_2 r_2 (\mathbf{g} - \mathbf{x}_i)$

    \Else
      \State $\mathbf{v}_i \gets \omega \mathbf{v}_i
      + c_2 r_2 (\mathbf{g} - \mathbf{x}_i)$

      \If{$k \ge N_e$ \textbf{and} $\mathrm{rand} < m_p$}
        \State $\mathbf{x}_i \gets \mathbf{g}(2\alpha\eta + 1 - \alpha)$
        \State \textbf{continue}
      \EndIf
    \EndIf

    \State $\mathbf{x}_i \gets \mathbf{x}_i + \mathbf{v}_i$

    \If{$F(\mathbf{x}_i) < F(\mathbf{p}_i)$}
      \State $\mathbf{p}_i \gets \mathbf{x}_i$
      \If{$F(\mathbf{p}_i) < F(\mathbf{g})$}
        \State $\mathbf{g} \gets \mathbf{p}_i$
      \EndIf
    \EndIf
  \EndFor

  \State Update inertia:
  \[
    \omega = \omega_{\max}
    - k \frac{\omega_{\max} - \omega_{\min}}{K_{\max}}
  \]

\EndFor

\end{algorithmic}
\end{algorithm}

The parameters of the algorithm are shown in Table~\ref{tab:PSO}.

\begin{table}
\caption{\label{tab:PSO} 
Hyperparameters of PSO-TPME used in this study.}
\begin{ruledtabular}
\begin{tabular}{lr}
Parameter description & Value \\ 
\hline
Number of unknown variables & 25 \\
Lower bound of decision variables & 0 \\
Upper bound of decision variables & 100 \\
Maximum velocity & 20 \\
Minimum velocity & -20 \\
Cognitive acceleration coefficient $c_1$ & 1.495 \\
Social acceleration coefficient $c_2$ & 1.495 \\
Maximum inertia weight $\omega_{\max}$ & 0.9 \\
Minimum inertia weight $\omega_{\min}$ & 0.1 \\
Max number of iterations & 200 \\
Population (swarm) size & 20 \\
Elitism start iteration $N_e$ & 2 \\
Mutation probability $m_p$ & 1\% \\
Mutation range $\alpha$ & 0.5 \\
Classification percentage $\rho$ & 2\% \\

\end{tabular}
\end{ruledtabular}
\end{table}

\section{Repeatability and Accuracy of Hot-Wire Anemometry}
\label{pta}

\begin{figure}[ht]
    \centering
    \includegraphics[width=0.5\textwidth]{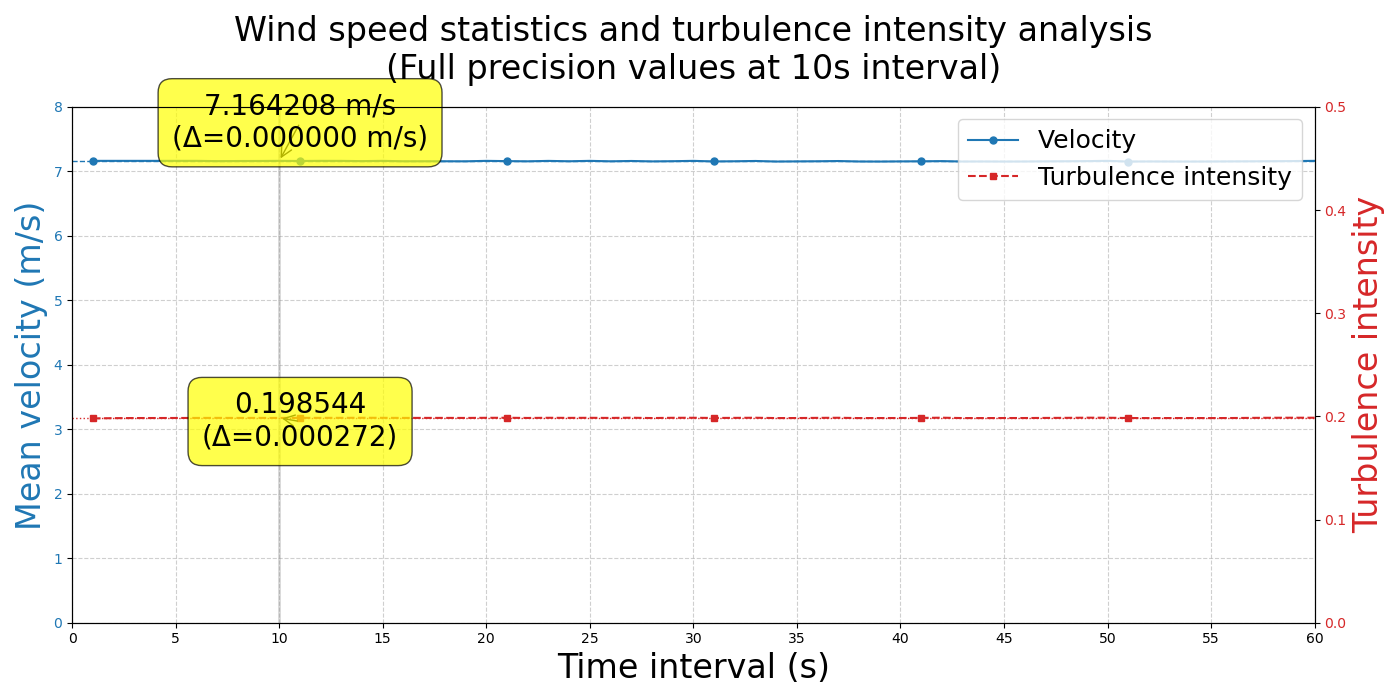}
    \caption{Wind Speed Measurement Stability Test}
    \label{fig:WSMST}
\end{figure}

During the experiment, the output power of the FAWG was adjusted to its maximum level and maintained for ten minutes to ensure the wind field reached a steady state. Subsequently, the wind speed measurement procedure was initiated, conducting real-time monitoring of wind speed for one continuous minute.
The experimental results are shown in FIG~\ref{fig:WSMST}, where the horizontal axis represents the sampling time interval (unit: $s$), and the vertical axis indicates the average wind speed (unit: $m/s$) and turbulence intensity for each time period. By comparing the measurements from the first 10 seconds with the 60-second long-term average, it was found that the error in average wind speed was zero, and the turbulence intensity error was merely 0.000272. These results fully satisfy the experimental accuracy requirements.
Based on this analysis, this study ultimately determined 10 seconds as the standard sampling time for the hot-wire anemometer. This selection not only ensures measurement data reliability but also significantly improves experimental efficiency.

Based on the data shown in the figure, this study concludes that the optimal duration for wind speed measurement is 10 seconds. At this point, the measured values show clear convergence, meaning that the wind speed readings tend to stabilize in consecutive measurements without significant fluctuations. This indicates that a 10-second measurement duration is sufficient to capture the representative state of the wind field, thereby ensuring the accuracy and reliability of the obtained data. All further measurements and data analysis are based on this optimal measurement duration.

In this experiment, the wind speed measurement program was first initiated to prepare the data acquisition environment. Subsequently, the FAWG device was activated to generate the required wind field conditions. The purpose of the experiment was to determine the time required for the wind field to reach a stable state. By precisely controlling the parameters of the FAWG and synchronizing the start of the measurement program, this study aims to record the entire process from the initialization of the wind field to its full stabilization. A detailed description of the experimental results is as follows:

\begin{figure}[hbt!]
    \centering
    \includegraphics[width=0.5\textwidth]{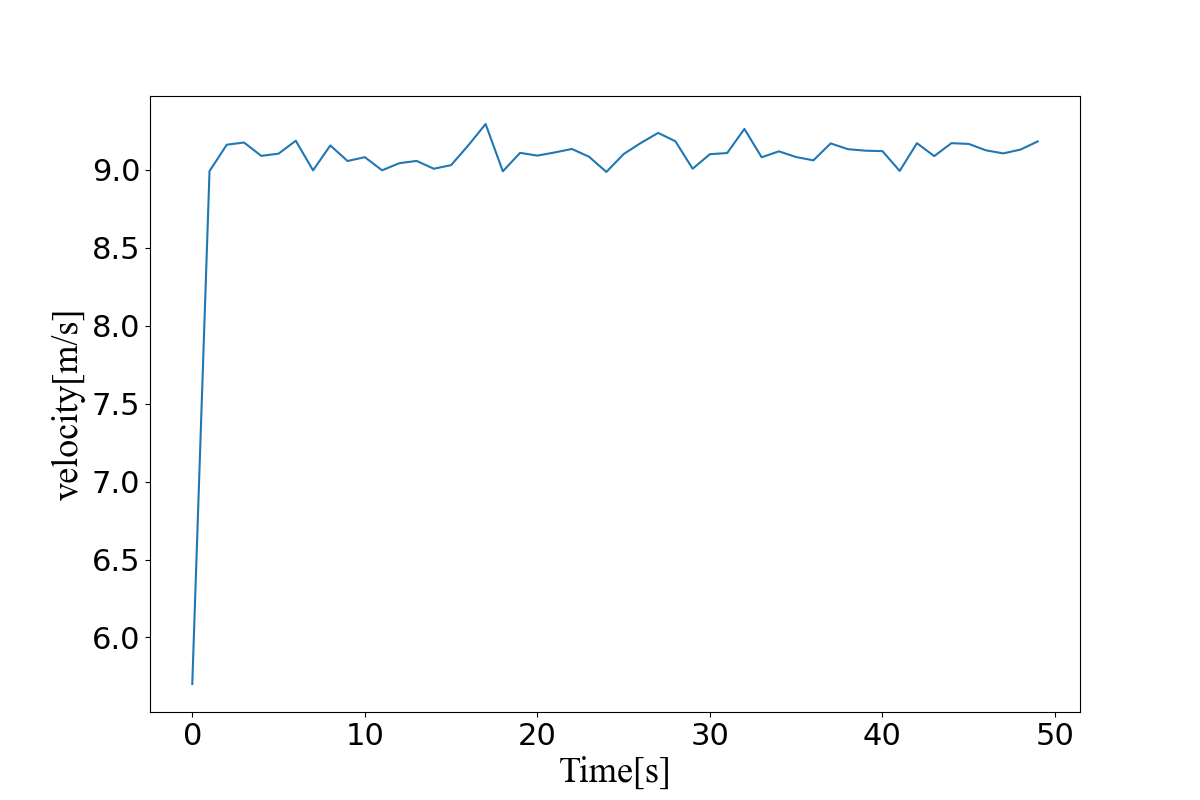}
    \caption{Wind Field Stabilization Measurement Result 1}
    \label{fig: WFSMR1}
\end{figure}

\begin{figure}[hbt!]
    \centering
    \includegraphics[width=0.5\textwidth]{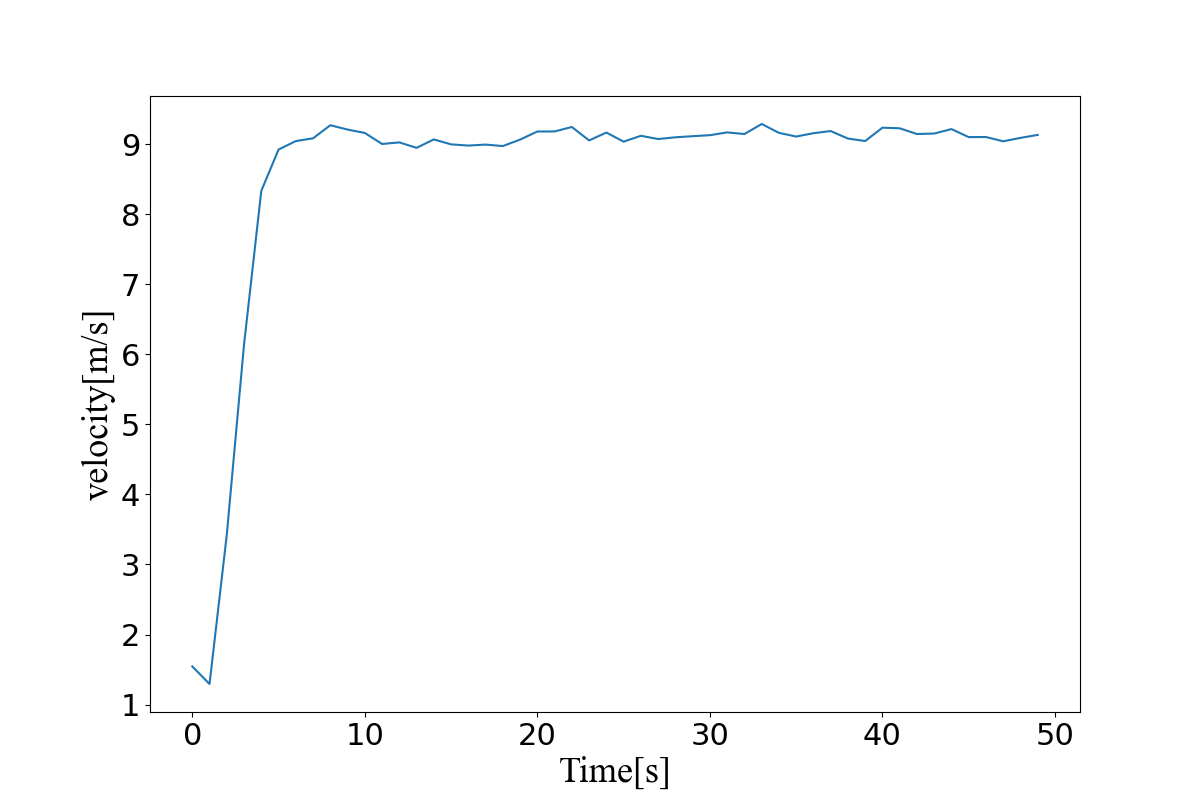}
    \caption{Wind Field Stabilization Measurement Result 2}
    \label{fig: WFSMR2}
\end{figure}

Based on the analysis of the data shown in the FIG~\ref{fig: WFSMR1}and~\ref{fig: WFSMR2}, the wind speed demonstrates significant stability after a 10-second measurement period. At this point, the wind speed value reaches a constant state with no significant changes occurring. This indicates that the wind field requires 10 seconds to achieve a dynamic equilibrium state after the FAWG device is activated or a new wind speed is set. Therefore, this study selects 10 seconds as the time threshold required for the wind field to stabilize after the FAWG is turned on.

\begin{acknowledgments}
This work is supported 
by the National Natural Science Foundation of China (NSFC) 
through grants 
 W2541002, 
 12572260,  
and 12172109, 
and by the Shenzhen Science and Technology Innovation Program 
under grants KJZD20230923115210021  
and JCYJ20220531095605012. 
We appreciate generous technical and scientific support from the HangHua company (Dalian, China).

We appreciate valuable stimulating discussions with
Stefano Discetti, Andrea Ianiro, 
Franz Raps, Jincai Yang  and Yang Yang.
We appreciate generous technical and scientific support from the HangHua Company (Dalian, China).
\end{acknowledgments}

\section*{Data Availability Statement}
The data that support the findings of this study are available from the first 
author upon reasonable request.

\bibliography{main}

\end{document}